\documentclass[
superscriptaddress,twocolumn,
preprintnumbers,
 amsmath,
 amssymb,
 prl,
]{revtex4-2}
\usepackage{amsfonts}
\usepackage{float}
\usepackage{placeins}
\usepackage{graphicx}
\usepackage{hyperref}
\usepackage{color,amsmath,amssymb,bm}
\usepackage{tensor}
\usepackage{lineno}

\usepackage[normalem]{ulem}  

\renewcommand{\sout}{\bgroup \color{red} \ULdepth=-.5ex \ULset}

\newcommand\sect[1]{\noindent \textbf{#1.}---}

\def \be {\begin{equation} }
\def \ee {\end{equation}}
\def \bes {\begin{subequations} }
\def \ees {\end{subequations}}

\def \a {\alpha}
\def \b {\beta}

\def \d {\delta}
\def \e {\epsilon}
\def \g {\gamma}

\def \o {\omega}

\def \l {\lambda}

\def \s {\sigma}

\def \vp {\bm{p}}

\def \vx {\bm{x}}
\def \vy {\bm{y}}
\def \vr {\bm{r}}
\def \vk {\bm{k}}
\def \G {\Gamma}

\def \<{\langle}
\def \>{\rangle}
\def \+{\dagger}
\def \le {\left}
\def \ri {\right}

\def \pd {\partial}
\def \baro {\mathfrak{w}}
\def \bark {\mathfrak{q}}

\def \sq {\mathfrak{q}}
\def \so {\mathfrak{w}}
\def \su {\mathfrak{u}}

\def \qperp
\def \CC {{\cal C}}

\newcommand{\bfk}{\mathbf{k}}

\begin{document}
\preprint{LA-UR-22-27822}
\title{
Does quark-gluon plasma feature an extended hydrodynamic regime?
}

\author{Weiyao Ke\footnote{weiyaoke@lanl.gov}}
\affiliation{Theoretical Division, Los Alamos National Laboratory, Los Alamos NM 87545, United States
}

\author{Yi Yin\footnote{yiyin@impcas.ac.cn}}
\affiliation{Quark Matter Research Center, Institute of Modern Physics, Chinese Academy of Sciences, Lanzhou 730000, China}

\date{\today}

\begin{abstract}
We investigate the response of the near-equilibrium quark-gluon plasma (QGP) to perturbation at non-hydrodynamic gradients. 
We propose a conceivable scenario under which sound mode continues to dominate the medium response in this regime. 
Such a scenario has been observed experimentally for various liquids and liquid metals. 
We further show this extended hydrodynamic regime (EHR) indeed exists for both the weakly-coupled kinetic equation in the relaxation time approximation (RTA) and the strongly-coupled ${\cal N}=4$ supersymmetric Yang-Mills (SYM) theory. 
We construct a simple but nontrivial extension of M{\"u}eller-Isareal-Stewart (MIS) theory, namely MIS*, and demonstrate that it describes EHR response for both RTA and SYM theory. 
This indicates that MIS* equations can potentially be employed to search for QGP EHR via heavy-ion collisions. 
\end{abstract}

\maketitle

\sect{Introduction}%
The properties of many-body QCD systems, particularly their behavior at different scales, have fascinated the high-energy nuclear physicists for decades. 
One prominent example of such a system is quark-gluon plasma (QGP) that is created in heavy-ion collision (HIC)~\cite{PHENIX:2004vcz}.
The remarkable and nontrivial agreement between hydrodynamic modeling and many results of HIC~\cite{Heinz:2013wva,Schenke:2021mxx} indicates that QGP, which is asymptotic free at short distances, behaves as a near-perfect fluid at long distances. 
However, much less is known about its properties at the intermediate scale where the characteristic momentum is too large for a fluid description and is too small to be treated perturbatively~\cite{Busza:2018rrf}.

One central question in this context is how an equilibrated QGP responds to an external disturbance with varying gradients. 
When the perturbation varies slowly in space and time, 
the hydrodynamic modes, such as sound and shear modes, dominate the response after a sufficiently long time.
In this hydrodynamic regime (HR), the medium's behavior is described by viscous hydrodynamics. 
The description beyond HR, in general, can be exceeding complex. 
It is commonly assumed that dynamics in such situations involve a tower of non-hydrodynamic excitations \cite{Kurkela:2019kip,Kurkela:2021ctp}.

Nevertheless, an alternative scenario may and does exist in condensed matter systems. 
Since the seminal measurement in Ref.~\cite{PhysRevLett.32.49}, there is mounting empirical evidence showing that a variety of liquids and liquid metals, such as liquid Zinc~\cite{PhysRevLett.114.187801}, Liquid K-Cs Alloys~\cite{PhysRevLett.85.5352}, can sustain sound modes extending from HR to wavelengths comparable to inter-atomic separations, see Refs.~\cite{RevModPhys.77.881,Trachenko_2015} for reviews. 
Though the physical origins of those so-called ``high-wavevector" sound modes are still under intensive investigation, 
they are found to be essential for understanding the properties of the material at non-hydrodynamic wavelength (e.g. Ref.~\cite{PhysRevLett.85.5352}).

The above observations exemplify the situation that
the damping rate of hydrodynamic excitations remains smaller than other excitations up to some critical wavenumber $q_{c}$ beyond HR.
In such an extended hydrodynamic regime (EHR), hydrodynamic modes still dominate the response at a time scale longer than the inverse of the gap (in damping rate) despite viscous hydrodynamics is not expected to describe their dispersion.
As we shall see explicitly below, EHR exists in representative microscopic theories such as the kinetic theory under relaxation time approximation (RTA) and the strongly coupled ${\cal N}=4$ super-symmetric Yang-Mills theory in large $N_{c}$ limit (SYM). 
This indicates the generality of EHR for gauge theories, regardless of the coupling strength.

In this letter, we propose EHR as a conceivable scenario for QGP.
In the context of the far-from-equilibrium stage of heavy-ion collisions, the extension of hydrodynamics in describing bulk evolution is under intensive study~\cite{Heller:2015dha,Romatschke:2017vte}. 
Little attention has been paid to the possibility that sound modes may dominate the near-equilibrium response at a significant gradient. 
What would we learn about QGP if such an EHR existed?
First, suppose so, the characterization of QGP at intermediate scales is simplified, that is, we can describe medium properties in EHR with a few parameters such as effective sound velocity and attenuation rate. 
Second, it will fill the gap in our knowledge about the emergence of QGP liquid from asymptotic parton gas.

Can we test the EHR scenario of QGP experimentally?
In HIC, the azimuthal asymmetries in the flow of produced hadrons $v_{n}$ have been measured and are commonly interpreted as the response to the initial eccentricity.
In smaller colliding systems, eccentricities are dominated by shorter scale fluctuations and the EHR response can be responsible for $v_{n}$ generation.
In fact, describing observed collectivity in those small colliding systems from non-hydrodynamic transport has already attracted considerable attentions~\cite{Kurkela:2019kip,Kurkela:2021ctp}. 
Besides, medium response to jet propagation can be another channel to probe the EHR behavior. 
In Refs. \cite{PhysRevC.95.044909,Chen:2017zte}, it was found that energy and momentum deposited by fast partons cause large gradient perturbations and the excited medium response is correlated non-trivially with high-energy jet production.

A phenomenological investigation of QGP EHR requires extending the standard hydrodynamic modeling of HIC to incorporate EHR response. 
To our knowledge, no serious attempt to construct such a model has been made to date.  We shall write down a simple but nontrivial extension of M{\"u}eller-Isareal-Stewart (MIS) theory, which we call MIS*, for this purpose.
We shall demonstrate that MIS* can successfully describe EHR response for different microscopically theories for the static and Bjorken expanding background.
We believe our construction represents an essential first step toward experimental search for EHR.

\sect{EHR in weakly- and strongly-coupled theories}%
In this section, we consider a conformal and uncharged system and focus on the response in the sound channel.

We begin with the RTA kinetic theory, where the distribution function $f(t,\vx,\vp)$ obeys 
\begin{align}
\label{RTA}
  p^{\mu}\pd_{\mu}f-\Gamma^{\l}_{\a\sigma}p^{\a}p^{\sigma}\frac{\pd f}{\pd p^{\l}} = -\frac{u\cdot p}{\tau_{R}}\, \le(
f - f_{{\rm eq}}
\ri)\, .
\end{align}
Here $\G^{\l}_{\a\b}$ denotes the metric connection and we use the ``mostly plus'' metric.  
The relaxation time $\tau_{R}$ controls the time scale at which $f$ approaches the equilibrium distribution $f_{{\rm eq}}=e^{\beta p\cdot u}$, where $\beta$ and $u^{\mu}$ denote inverse temperature and fluid velocity, respectively. 
For a conformal liquid, $\tau_R\propto \e^{-1/4}$ with $\e$ being the local energy density.
As shown in Ref.~\cite{Romatschke:2015gic},  the dispersion of the sound mode can be obtained from solving
\begin{align}
\label{sound-kin}
 (\sq^2+i C_{\pi}\so)+\frac{i C_{\pi}}{2 \sq}\le[\sq^2+3\so(iC_{\pi}+\so)L\ri]=0\, .
\end{align}
Here we use the dimensionless frequency $\baro\equiv \o\nu$ and wave-vector $\sq=\nu q$ where $\nu=4\eta_{0}/3(\e_{0}+p_{0})$, with $\eta_{0}$ the shear viscosity, $\e_{0}$ the energy density and $p_{0}$ the pressure of the background. $L=\ln\le(\frac{\baro-\bark+iC_{\pi}}{\baro+\bark+iC_{\pi}}\ri)$ and
the dimensionless ratio $C_{\pi}\equiv \tau_{\pi}/\nu$ is related to the shear relaxation time $\tau_{\pi}$.
For RTA kinetic theory, $\nu=4\tau_{R}/15$ and $C_{\pi}=15/4$~\cite{Romatschke:2015gic} while for SYM, $\nu=\beta/3\pi, C_{\pi}=3(2-\log 2)/2$~\cite{Baier:2007ix}.

Turning to SYM,  we employ the AdS/CFT correspondence which maps the correlator in the quantum field theory in $d$-dimensional space-time into a classical general relativity calculation in $d+1$ dimension. 
The excitations in sound channel can be found by solving~\cite{Kovtun:2005ev,Amado:2008ji} 
\begin{widetext}
\begin{align}
\label{Z-eq}
Z''-\frac{3\baro^{2}(1+\su^{2})+\bark^{2}(2\su^{2}-3\su^{4}-3)}{\su (1-\su^{2})(3\baro^{2}+\bark^{2}(\su^{2}-3))}\,Z'+\frac{4}{9}\times\frac{3\baro^{4}+\bark^{4}(3-4\su^{2}+\su^{4})+\bark^{2}(9\su^{5}-9\su^{3}+4\su^{2}\baro^{2}-6\baro^{2})}{\su (1-\su^{2})^{2}(3\baro^{2}+\bark^{2}(\su^{2}-3))}Z=0\, ,
\end{align}
\end{widetext}
where $Z(\su)$ depends on radial coordinate $\su$ of the extra dimension ranging from the boundary $\su=0$ to the horizon $\su=1$.
 Eq.~\eqref{Z-eq} has to be solved with the boundary condition $Z\sim (1-\su)^{-i\baro/3}$ as $\su\to 1$, corresponding to the in-falling wave at the horizon. 
Near the boundary $\su=0$, the solution can be written as $  Z(\su)\sim {\cal A}(\bark,\baro)(1+\ldots)+{\cal B}(\bark,\baro)\su^{2}+\ldots$
where the dots denote higher powers of $\su$, and ${\cal A},{\cal B}$ depend on $\baro, \bark$. 
The Dirichlet boundary condition ${\cal A}(\bark,\baro)=0$ gives the dispersion relation of excitations which we obtained numerically by using the Mathematica notebook that was written and made public by Jansen~\cite{Jansen:2017oag}.

\begin{figure*}

\includegraphics[width=.45\textwidth]{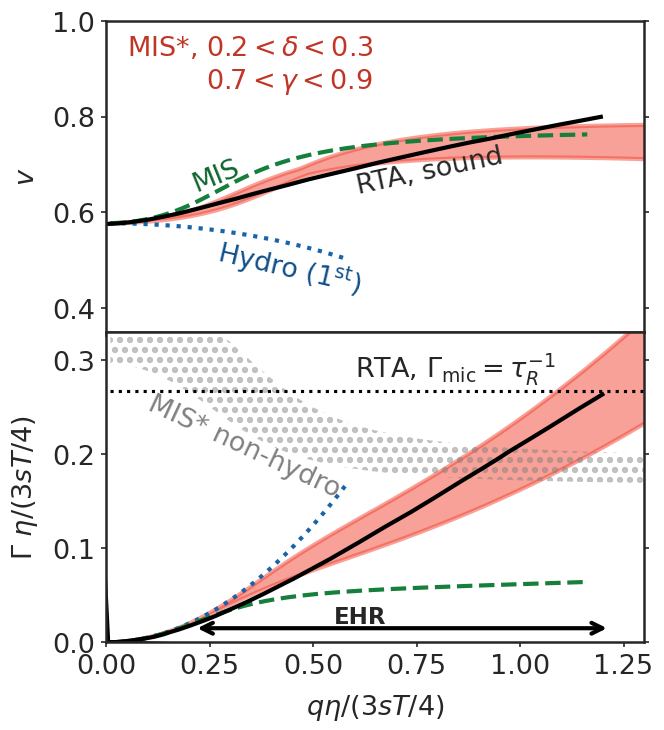}
\includegraphics[width=.45\textwidth]{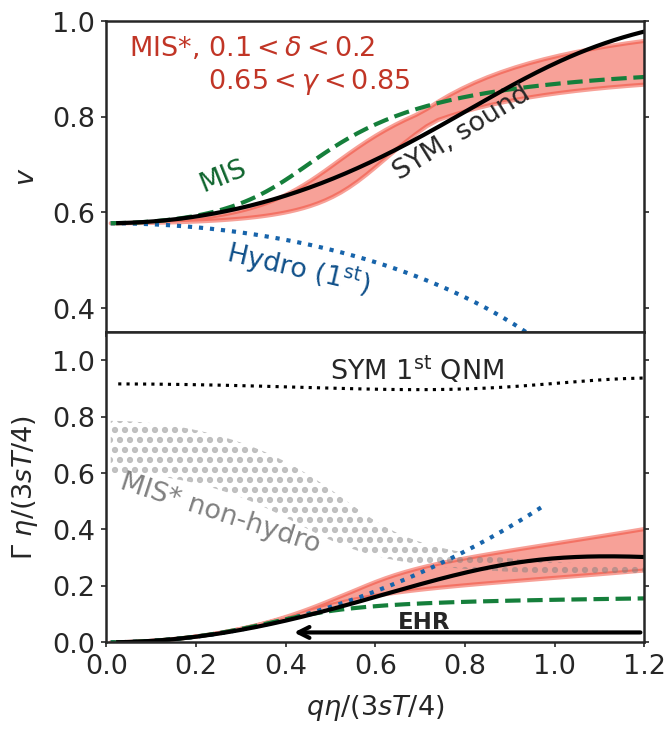}
\caption{
\label{fig:MIS-dis}
Sound phase velocity (the upper panel) and damping rate $\Gamma$ (the lower panel) as a function of gradient $\sq$ for RTA kinetic theory (left) and ${\cal N}=4$ SYM in strong coupling limit. 
Both $\Gamma$ and $\sq$ are scaled by $\eta/(3 sT/4)$. 
The damping rate of non-hydrodynamic excitations are shown in black dotted curves. 
The gap between those excitations and sound modes at non-hydrodynamic gradient indicates the existence of extended hydrodynamic regime (EHR). 
The first-order hydrodynamic and MIS results, shown in blue dotted and green dashed curves, respectively, fail to capture the key features of sound modes in EHR. 
In contrast, 
for a given range of model parameters (in red bands), MIS* efficiently describes EHR sound propagation. 
}
\end{figure*}

In Fig.~\ref{fig:MIS-dis}, 
we present the sound dispersion relation $\so_{\pm}(\sq)=\pm v(\sq)\sq- i \Gamma(\sq)$ in RTA kinetic theory (left) and SYM (right).
The phase velocity $v$ and sound attenuation rate $\Gamma$ (rescaled by $\nu$) are shown in red solid curves in the upper and lower panel, respectively.
Also shown are representative non-hydrodyamic excitations in black dashed curve. 
For comparison, 
we consider dispersion of excitations in sound channel in MIS theory~\cite{Romatschke:2009im,Hong:2010at}
\begin{align}
  \label{sound-MIS}
\baro^{2}-c^2_s\,\bark^2+ \frac{i\baro \bark^2}{(1-i C_\pi \baro)} &=0\, ,
\end{align}
where sound velocity $c^{2}_{s}=1/3$ for a conformal fluid. 
The solutions to Eq.~\eqref{sound-MIS} includes a pair of sound modes and a purely dissipative mode at finite $C_{\pi}$.
In the limit $C_{\pi}\rightarrow 0$, Eq.~\eqref{sound-MIS} reduces to the familiar expression in first-order hydrodynamic
$ \baro^{2}-c^2_s\,\bark^2+ i\baro \bark^2=0$.
Sound dispersion in first-order hydrodynamics and MIS, 
obtained by solving \eqref{sound-MIS} with $C_{\pi}=0$ and fixed $C_{\pi}$, 
are shown in blue and magenta curves, respectively, in Fig.~\ref{fig:MIS-dis}. 
Plotting them at non-hydrodynamic gradient will tell us the regime where conventional hydrodynamic theories cease to be a good description of the response.

For RTA kinetic theory, 
the sound modes are gapped from the non-hydrodynamic excitations, i.e. quasi-particles with damping rate $\tau^{-1}_{R}$, up to a critical value $\sq_{c}=1.2$~\cite{Romatschke:2015gic}. 
Far below $\sq_{c}$, the sizable difference from first-order hydrodynamics is seen in damping rate and is even more so in phase velocity even for $\sq>0.2$.  
Therefore, $0.2<\sq<1.2$ may be viewed as EHR for RTA kinetic theory, 
noting that EHR does not have a sharp boundary and hence the values quoted here are only for illustrative purpose.

Turning to SYM, we notice that the gap between sound modes and non-hydrodynamic excitations, which are a tower of quasi-normal modes, remains open for any $\sq$ under study (see also Refs.~\cite{Amado:2008ji,Fuini:2016qsc}. 
Indeed, the latter found that the gap only vanishes as $q^{-1/3}$ at asymptotic large $\sq$).
The first-order hydrodynamics fails to describe dispersion, notably for the phase velocity, for $\sq > 0.3$. 
So we conclude that the EHR also exists in SYM and corresponds to $\sq>0.3$.  
Besides, EHR can be identified in other strongly coupled theories, e.g. Ref.~\cite{Edalati:2010pn}.
We stress that we analyze the near-equilibrium gradient expansion, rather than the far-from-equilibrium bulk evolution. 
For the latter case, the hydrodynamic description appears to be unexpectedly reasonable even when the gradients are large in the same SYM theory~\cite{Chesler:2009cy,Heller:2011ju,Chesler:2015bba,Casalderrey-Solana:2013aba,Casalderrey-Solana:2013sxa}.

We summarize features that kinetic and SYM theory share in the EHR, which, interestingly, are also seen in various materials mentioned in the Introduction.
First, the phase velocity, or the ``effective stiffness'', becomes supersonic, $v\geq 1/3$. 
Second, the damping rate $\Gamma(q)$ is always smaller than that of the first-order hydrodynamics at fixed $\sq$. 
Those behaviors may be anticipated on physical grounds as follows.  
As gradient $\sq$ grows, more and more d.o.fs fall out of equilibrium and can not respond to the compression, resulting in larger stiffness i.e. a larger effective sound velocity.
At the same time, those off-equilibrium d.o.fs would not contribute to the dissipative process, leading to a smaller damping rate.
The above picture is admittedly speculative, but it illustrates that studying sound propagation can be instrumental in characterizing the medium properties in EHR.

\sect{Dynamical Models for EHR}
As noted in the Introduction, building dynamical models with EHR response is necessary for searching for EHR through data-modeling comparison.  
The commonly used MIS theory or its variant, which includes corrections second order in gradient, is not suitable for the present purpose. 
Indeed, as already shown in Fig.~\ref{fig:MIS-dis}, 
MIS results only give a modest improvement as compared with first-order hydrodynamics but generically under-estimates the attenuation in EHR.
One may consider adding higher order gradient terms, see Refs.~\cite{Grozdanov:2019kge,Grozdanov:2019uhi,Heller:2020uuy,Heller:2020hnq} for recent development on the convergence of gradient expansion (see also Refs.~\cite{Lublinsky:2009kv,Heller:2014wfa,Heller:2021yjh} for different attempts to improve hydrodynamics).
Although systematic in principle, doing so would result in the proliferation of input parameters.

Instead, we aim at constructing a model such that a) it reduces to viscous hydrodynamics in long-wavelength limit; 
b) for a given microscopic theory, it could describe sound propagation in EHR (or a least part of it) with a minimum number of model parameters.
Below, we propose an extension of the MIS equation, MIS*, containing two key additional model parameters as compared with first-order hydrodynamics and demonstrate it serves the purpose. 
Our construction is partly inspired by the Hydro+ framework~\cite{Stephanov:2017ghc}.

\sect{The construction of MIS*}
Consider the decomposition of the stress-energy tensor $T^{\mu\nu}= T^{\mu\nu}_{(0)}+\pi^{\mu\nu}$, where $T^{\mu\nu}_{(0)}=\e\, u^{\mu}u^{\nu} +p \Delta^{\mu\nu}$ denotes the the ideal fluid energy-momentum tensor with $ \Delta^{\mu\nu}= g^{\mu\nu}+u^{\mu}u^{\mu}$. 
In viscous hydrodynamics, the non-equilibrium corrections to energy-momentum tensor, $\pi^{\mu\nu}$, becomes $\pi^{\mu\nu}_{(1)}=\eta \s^{\mu\nu}$, where the shear strength is given by $\sigma^{\mu\nu}=\pd^{\mu}_{\perp}u^{\nu}+\pd^{\mu}_{\perp}u^{\nu}-(2/3)\Delta^{\mu\nu} \theta$ with $\theta=\pd\cdot u,
\pd^{\mu}_{\perp}\equiv \Delta^{\mu\nu}\pd_{\nu}$. 
The MIS theory treats $\pi^{\mu\nu}$ as a dynamical variable which relaxes to $\pi^{\mu\nu}_{(1)}$ at time scale $\tau_{\pi}$. 
However, in EHR where the time scale can be shorter than $\tau_{\pi}$, $\pi^{\mu\nu}$ fall out-of-equilibrium and does not contribute to the dissipation. 
As a result, EHR sound is under-damped in MIS. 
To improve the description of sound propagation, in MIS*, we further divide $\pi^{\mu\nu}$ into two parts
\begin{align}
  \pi^{\mu\nu}=
  \pi^{\mu\nu}_{s}+\pi^{\mu\nu}_{f}\, , 
\end{align}
and evolve $\pi_{s},\pi_{f}$ at different relaxation times $\tau'_{\pi}, \tau''_{\pi}$, respectively.
By design, we require $\tau'_{\pi}\gg \tau''_{\pi}$ such that in EHR regime, the typical time scale is comparable or shorter than $\tau'_{\pi}$ but is much longer than $\tau''_{\pi}$. 
Consequently, in EHR,  $\pi^{\mu\nu}_{f}$ should approach a fixed form which we take to be $\eta' \sigma^{\mu\nu}$.
Here $\eta'<\eta$ controls the effective viscosity. 
Explicitly, we propose the following equations
 \begin{align}
\label{Pi-s}
\mathcal{D}\pi^{\mu\nu}_{s} &= -\frac{\pi^{\mu\nu}_{s}+(\eta-\eta')\sigma^{\mu\nu}}{\tau'_{\pi}}
+R \, ,
\\
\label{Pi-f}
\mathcal{D}\pi^{\mu\nu}_{f} &= -\frac{\pi^{\mu\nu}_{f}+\eta'\sigma^{\mu\nu}}{\tau''_{\pi}} \, , 
\end{align}
where $D=u\cdot \pd$ and $R=-\frac{4}{3}\pi^{\mu\nu}_{s}\theta+\ldots$ denotes other possible second order gradient terms which do not contribute to the sound dispersion. 
In practice, we may fix the form of $R$ by requiring Eq.~\eqref{Pi-s} to become usual BRSSS second order hydrodynamics~\cite{Baier:2007ix} or its variants in the limit $\eta'=0$.

Eqs.~\eqref{Pi-s}, \eqref{Pi-f} together with $\pd_{\mu}T^{\mu\nu}=0$ constitute MIS* equations, and are one of the main results in this letter. 
In the time scale longer than $\tau'_{\pi}$, $\pi^{\mu\nu}_{s}\to (\eta-\eta')\s^{\mu\nu}, \pi^{\mu\nu}_{f}\to \eta'\s^{\mu\nu}, R\to 0$ and hence MIS* reduces to the first-order fluid dynamics. 
If we take the limit $\tau''_{\pi}\to 0$, $\pi^{\mu\nu}_{f}$ becomes $\eta'\sigma^{\mu\nu}$ and is no longer dynamical. 
This is the limit we shall use below for illustration, though in a realistic simulation of MIS*, a finite $\tau''_{\pi}$ is needed to ensure causality.
We expect that the dynamics in EHR should be sensitive to $\eta',\tau'_{\pi}$ rather than $\tau''_{\pi}$ as far as $\tau'_{\pi}\gg \tau''_{\pi}$.

\sect{Sound mode in MIS*}%
The excitations in the sound channel can be obtained by linearizing MIS* equations around a homogeneous and static background.
In the limit $\tau''_{\pi}/\tau'_{\pi}\to 0$, we have
\begin{align}
\label{KY-sound}
    \baro^{2}-c^2_s\,\bark^2+  i\le[\d+\frac{(1-\d)}{(1-i \gamma C_\pi \baro)}\ri]\baro \bark^2 &=0\, ,
\end{align}
where we have defined two dimensionless ratio $(\d,\g)=(\eta'/\eta, \tau'_{\pi}/\tau_{\pi})$.
Note for $\delta=0$ and $\g=1$, Eq.~\eqref{KY-sound} reduces to Eq.~\eqref{sound-MIS}.
Similar to MIS theory, the excitations of MIS* from Eq.~\eqref{KY-sound} includes a pair of sound modes and a dissipative mode. 
The boundary of EHR in MIS* is determined by the ``level-crossing point'', $q'_{c}$ where the sound damping rate equals to that of dissipative mode.
In Fig.~\ref{fig:parameter} in the Appendix.~\ref{sec:para}, we show the dependence of dispersion on $\d, \g$. 
Thereby, we demonstrate that by tuning them, MIS* has the flexibility and capability of describing a class of sound propagation up to $q'_{c}$.

The comparison between sound dispersion in MIS* with that in RTA and SYM theory, as shown in Fig.~\ref{fig:MIS-dis}, is encouraging. 
For the $0.2<\d<0.3, 0.7<\g<0.9$, we find MIS* describes RTA sound mode well up to $\sq$ values around $\sq'_{c}=0.95$. 
For SYM system, a different range $0.1<\d<0.2, 0.65<\g<0.8.5$ provides a reasonable description up to $\sq'_{c}=0.7$. 
Given its simplicity, we do not anticipate the MIS* covering the full EHR in the two microscopic theories. 
Notwithstanding, MIS* extends the description from HR to a significant part of EHR in both cases.

%
%
%
%
\begin{figure*}
 \includegraphics[width=.9\textwidth]{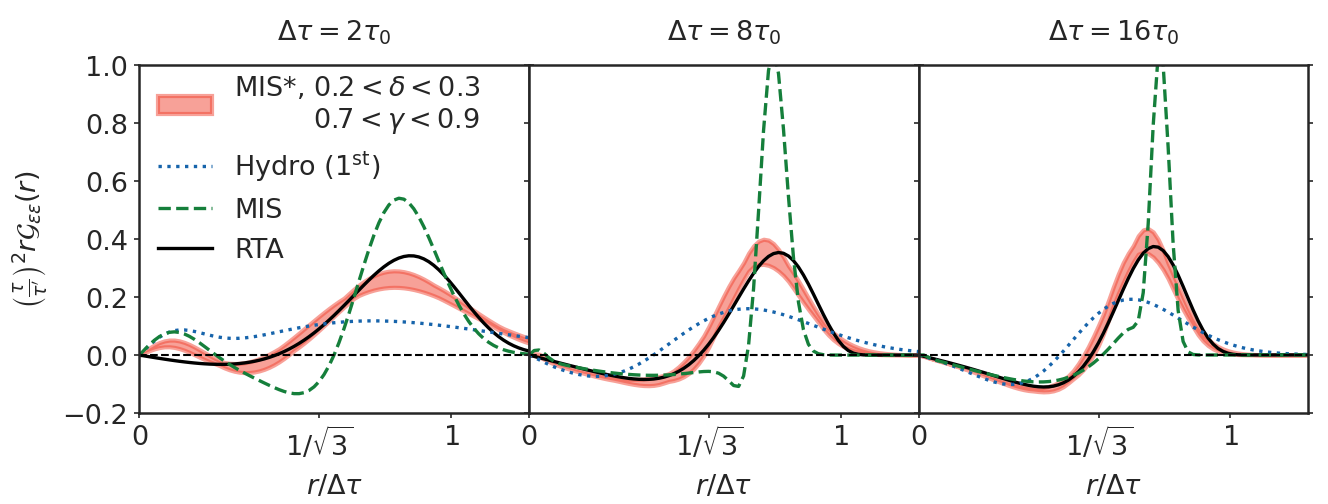}
\caption{
\label{fig:ee}
Energy-energy response function~\eqref{Gee-def} (rescaled by the appropriate power of $\tau/\tau'$)  as a function of $r/\Delta \tau$ ($\Delta\tau=\tau-\tau'$)  for the Bjorken expanding plasma. 
RTA kinetic theory, first order hydrodynamics and MIS theory results are shown in the solid, blue dotted and green dashed curves, respectively. 
The response function of MIS*, computed from the same range of model parameters $\delta, \g$ as used in Fig.~\ref{fig:MIS-dis} (left) to describe EHR sound dispersion, are plotted in the red band. 
Three values of $\tau$ are chosen to represent the response function at early, intermediate and late time (from left to right) with fixed $\tau'=2 \tau_{0}$, see text.
In hydrodynamic limit, the peak location of response function should approach $c_{s}=1/\sqrt{3}$. 
}
\end{figure*}

\sect{Bjorken expansion}%
One advantage of MIS* is that once its model parameters are fixed by matching to sound dispersion, it can readily be applied to expanding background.
To demonstrate this, we consider linearized response for a boost-invariant Bjorken expanding background specified by the evolution of $\e_{0}(\tau)$ vs. Bjorken time $\tau$.
For simplicity, we concentrate on the case that the perturbation only depends on $\vy$, the spatial vector lying entirely in the plane transverse to $z$-direction (longitudinal direction).
The response of $T^{\mu\nu}$ to external disturbance is characterized by several independent response functions~\cite{Kurkela:2018vqr}, see also Appendix.~\ref{fig:G-all}. 
Below, we focus on the energy-energy response function ${\cal G}_{\e\e}$ that evolves the initial energy perturbation $\delta \e(\tau',\vy')$ to
\begin{align}
\label{Gee-def}
  \delta \e(\tau,\vy)=\int d^{2}\vy'\, {\cal G}_{\e\e}(r;\tau,\tau')\, \delta\e(\tau',\vy')\, .
\end{align}
where $\vr\equiv \vy-\vy'$.
To obtain ${\cal G}_{\e\e}$ in RTA equation, we expand the distribution function $f(\tau,\vy;\vp)=f_{0}(\tau;\vp)+\delta f(\tau,\vy;\vp)$.
We first determined the background solution $f_{0}(\tau;\vp)$ with a relaxation time $\tau_R(\tau)={\e(\tau)}^{-1/4}$ starting from $\tau_0=\tau_R(\tau_0)$.
Then the linearized RTA equation.~\eqref{RTA} is solved numerically to obtain ${\cal G}_{\e\e}(\bfk;\tau,\tau')$ in Fourier space. 
When transforming it back to the $\vy$ space, a smearing function $\exp\left\{-\bfk^2/[2(4/\tau_{0})^2]\right\}$ is applied to regulate the integration at large $k$.

In Fig.~\ref{fig:ee}, we take $\tau'=2\tau_{0}$,
meaning we are considering the near-equilibrium expanding plasma rather than the far-from-equilibrium one~\cite{Kamata:2020mka}, 
and show ${\cal G}_{\e\e}$ at three representative times $\Delta\tau =2,8,16\tau_{0}$ as a function of $r/\Delta\tau$ where $\Delta\tau=\tau-\tau'$.
The RTA response function are then compared to first-order hydrodynamics ($\pd_{\mu}T^{\mu\nu}$ with $\pi^{\mu\nu}=\eta \sigma^{\mu\nu}$), MIS and MIS* theory, 
see our companion paper for more details. 
The hydrodynamic curves match with the kinetic ones at a very late time, say $\tau=16\tau_{0}$, but their differences are significant at those earlier times. 
In particular, the peak of $G_{\e\e}$,$(r/\Delta \tau)_{\rm peak}$, approaches $c_{s}=\sqrt{1/3}$ as $\tau$ increases, but is always larger than $c_{s}$, in accordance with supersonic nature of EHR sound propagation. 
Rather than improving the description at early times, the MIS theory introduces spurious shocks, see also Ref.~\cite{Hong:2011bd}. 
This can be understood by returning to Fig.~\ref{fig:MIS-dis} where we see MIS underestimates the sound attenuation in EHR.
Remarkably, MIS* response function, determined with the same range of parameters $\delta,\gamma$ as used to mimic EHR sound dispersion (see Fig.~\ref{fig:MIS-dis}), agree with RTA response from early to late times.
Equally impressive agreement is seen for the four other energy-momentum response functions,
see Fig.~\ref{fig:G-all} in Appendix.~\ref{sec:G}.
This convincingly indicates that describing EHR sound propagation is key to characterize response beyond conventional hydrodynamic regime, and MIS* serves that purpose.

\sect{Summary}%
To summarize,
we consider the Extended Hydrodynamic Regime (EHR) scenario for QGP, where sound modes are gapped from other excitations at a non-hydrodynamic gradient. 
We illustrate the existence of EHR in RTA kinetic theory and strongly coupled ${\cal N}=4$ SYM.
Our study is partly inspired by the observation of EHR in a class of liquids and liquid metals.

In the view that EHR scenarios should be explored experimentally in HIC, 
we construct hydrodynamic-like equations, MIS*, and demonstrate that it describes sound propagation in RTA kinetic theory and strongly coupled SYM theory in EHR. 
We also show the success of MIS* in describing RTA response in a Bjorken expanding medium. 
They together serve as proof of the principle that MIS*, with suitable refinement, can be implemented in future quantitative studies of small colliding systems~\cite{Nagle:2018nvi} and jet-medium interaction~\cite{PhysRevC.95.044909,Chen:2017zte}. 
Doing so allows us to identify observables sensitive to EHR and subsequently gain crucial lessons about intermediate-scale QGP through future data-model comparison.

\begin{acknowledgments}
We thank Xiaojian Du,  Shu Lin, Krishna Rajagopal, Li Yan for valuable discussions and Michal Heller, Soeren Schlichting for helpful comments on the draft. 
This work is supported by the US Department of Energy through the Office of Nuclear Physics and the LDRD program at Los Alamos National Laboratory (WK). Los Alamos National Laboratory is operated by Triad National Security, LLC, for the National Nuclear Security Administration of U.S. Department of Energy (Contract No. 89233218CNA000001). This work is also supported by the Strategic Priority Research Program of Chinese Academy of Sciences, Grant No. XDB34000000 (YY).
\end{acknowledgments}

\begin{appendix}

\section{The model parameter dependence of MIS* dispersion relation
\label{sec:para}
}

\begin{figure*}
    \centering
    \includegraphics[width=0.45\textwidth]{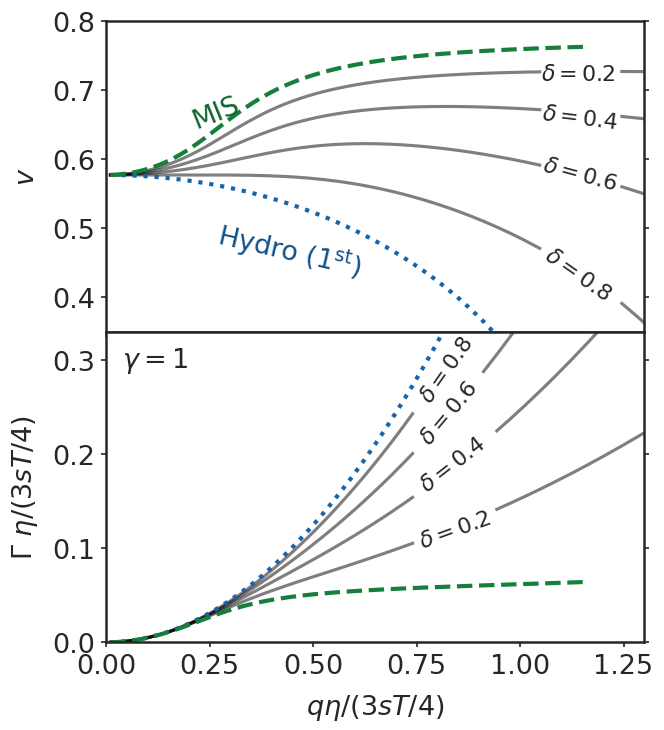}
    \includegraphics[width=0.45\textwidth]{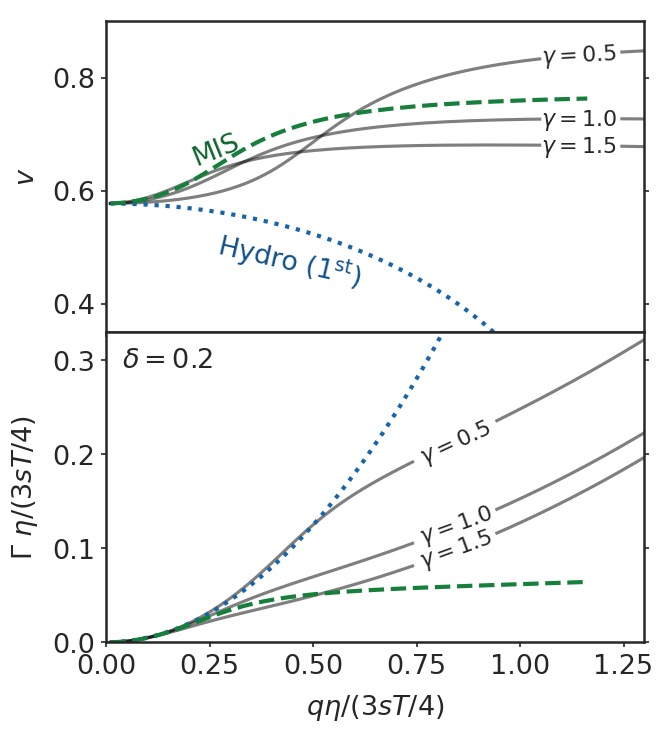}
    \caption{
    \label{fig:parameter}
  The sound velocity (the upper panel) and damping rate (the lower panel) as functions of rescaled momentum $\sq$ with different choices of $\delta,\gamma$ for MIS* equations. 
  The results are obtained by solving Eq.~\eqref{KY-sound}.  
  (Left):  $\gamma=1$ with $\delta=0.2, 0.4, 0.6, 0.8$;
  (right): $\delta=0.2$ with $\gamma=0.5, 1.0, 1.5$.
  }
\end{figure*}

We here illustrate the dependence of sound dispersion relation from MIS* equations, constructed in the main text, on the two essential model parameters $\delta, r$ by solving Eq.~\eqref{KY-sound} numerically. 
In Fig.~\ref{fig:parameter} (left), 
we present results with $\gamma=1$ but varying $\delta$. 
Note, in this case, MIS* reduces to the first order hydrodynamics for $\delta=0$ and to MIS for $\delta=1$. 
In Fig.~\ref{fig:parameter} (right),
we further show the dependence of the dispersion on $\g$ with fixed $\delta=0.2$. 
As we anticipated early, a larger $\g$ implies a greater attenuation rate. 
We observe while MIS* results agree with first order hydrodynamics and MIS in hydrodynamic regime as they should, 
the sound propagation and attenuation are sensitive to the value of $\delta,\g$.

\section{Energy-momentum response functions
\label{sec:G}
}
\begin{figure*}
\includegraphics[width=.9\textwidth]{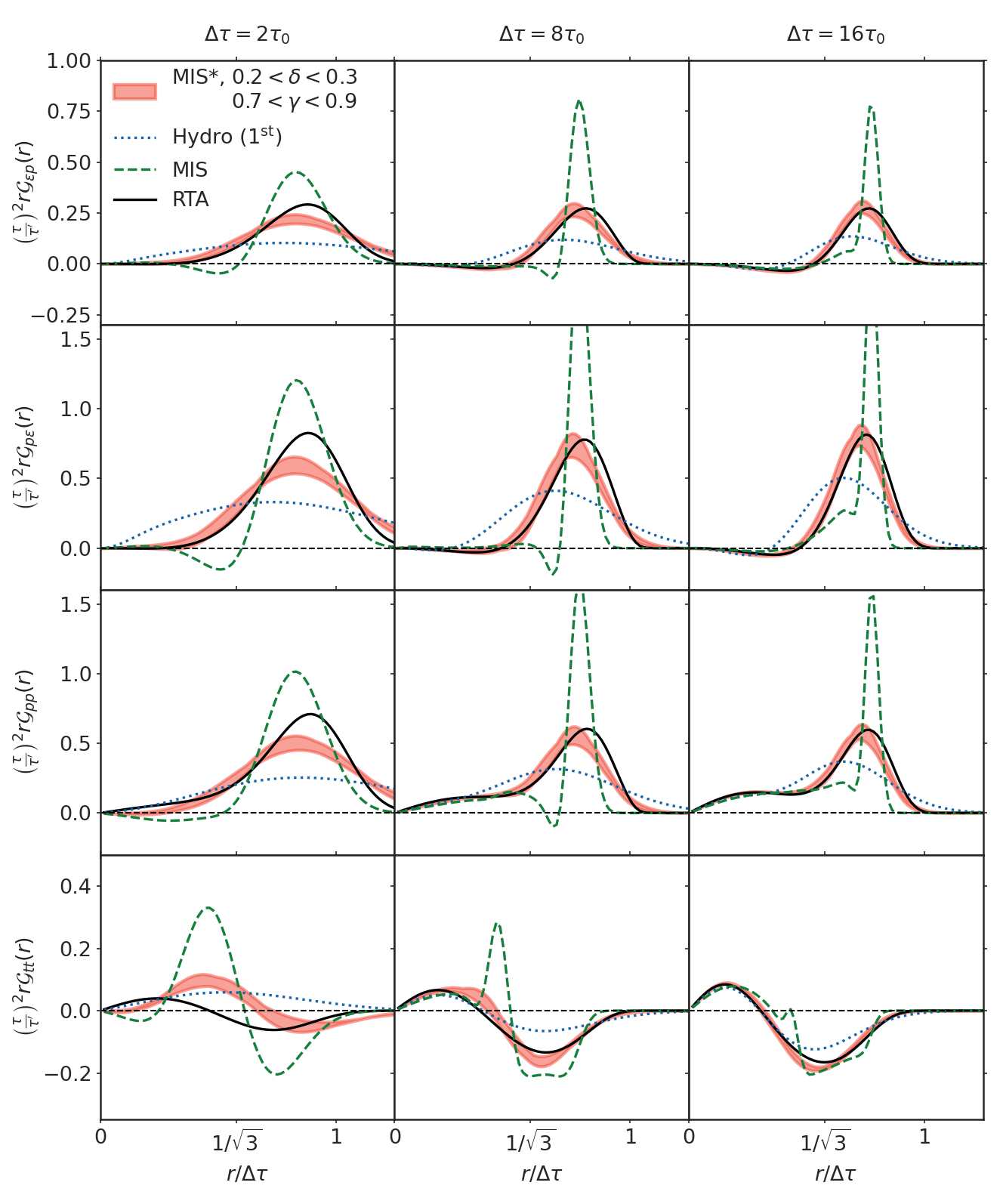}
\caption{
\label{fig:G-all}
The same as Fig.~\ref{fig:ee}, but for the four other independent energy-momentum response functions defined in Eq.~\eqref{G-list}.
}
\end{figure*}

Consider linearized energy and momentum propagation on top of a background Bjorken expanding plasma. 
The stress-energy tensor can be split into a background part and a perturbation part
\begin{align}
    T^{\mu\nu}=T^{\mu\nu}_{0}(\tau)+\delta T^{\mu\nu}(\tau,\vy)\, ,
\end{align}
where $T^{\mu\nu}_{0}(\tau)$ only depends on the Bjorken proper time $\tau$ and is diagonalized. 
We shall focus on the case that the perturbation only depends on $\vy$, the spatial vector lying entirely in the plane transverse to $z$-direction.
For given initial perturbation $\delta T^{\mu\nu}(\tau_{0}, \vy_{0})$ at $(\tau', \vy')$, 
we, following Ref.~\cite{Kurkela:2018vqr}, define the response function
\begin{align}
    \delta T^{\mu\nu}(\tau,\vy)= \int d^{2}\vy'\, G^{\mu\nu}_{\a\b}(\vy-\vy';\tau,\tau')\, T^{\a\b}(\vy',\tau')\, .
\end{align}
where we have used fact that the response function depend only on the different $\vr=\vy-\vy'$. 
If we concentrate on the energy-momentum response to the initial energy density $\delta T^{\tau\tau}$ and initial transverse momentum density  $\delta T^{0 i}$ ($i=x,y$),  there are five independent components which can be defined through the relation (Here and hereafter, we suppress the dependence on $\tau,\tau'$)
\bes
\label{G-list}
\begin{align}
&\,    G^{\tau\tau}_{\tau\tau}(r)\equiv {\cal G}_{\epsilon\epsilon}(r)\, , 
\\
&\,
G^{\tau i}_{\tau\tau}(\vr)\equiv \hat{r}^{i}\, {\cal G}_{\e p}(r)
    \qquad
  G^{\tau\tau}_{\tau i}(\vr)\equiv\hat{r}^{i}\, {\cal G}_{p \e}(r)\, ,
    \\
&\,  G^{\tau i}_{\tau j}(\vr)= \hat{r}^{i}\hat{r}^{j}{\cal G}_{pp}(r)+\le(\delta^{ij}- \hat{r}^{i}\hat{r}^{j}\ri){\cal G}_{tt}(r)\, . 
\end{align}
\ees
Physically, $G^{\epsilon}_{p}$ describes energy density at $\tau$ induced by the momentum density disturbance projected along direction $\hat{r}$ at initial time $\tau'$ and other response functions can be interpreted in a similar way. 
We are interested in the response of the plasma near equilibrium, so we take $\tau_{1}>\tau_{0}$ where $\tau_{0}$ denotes the thermalization time. 
In practice, we first compute $G^{\mu\nu}_{\a\b}$ in Fourier space $\vk=(k_{x},k_{y})$ by solving linearized RTA kinetic equation and then obtain response functions in real space with an appropriate Fourier transformation.

We have conducted detailed numerical tests of MIS* by further computing those response functions for the first-order hydrodynamics, MIS and MIS* that we are proposing.
See our upcoming publications for further details. 
The results for $G_{\e p}, G_{p \e}, G_{pp},G_{tt}$ are shown in Fig.~\ref{fig:G-all}.
They, together with the plot for $G_{\e\e}$ given in the main text,
demonstrate that there is indeed a large space-time domain where MIS* works.

\end{appendix}

\bibliography{ref}

\end{document}